\documentclass[aps,twocolumn,prd,floats,nofootinbib,showpacs]{revtex4}

\usepackage{latexsym}
\usepackage{amssymb}
\usepackage[dvips]{graphicx}
\usepackage[applemac]{inputenc}
\usepackage[nottoc]{tocbibind} 
\usepackage{float}
\usepackage{fancyhdr}
\usepackage{amsfonts}
\usepackage{amsmath}
\usepackage{color}
\usepackage{mathrsfs}
\usepackage{bm}
\usepackage{epsfig}
\def\wtil{\widetilde}

\def\lam{\lambda}
\def\kap{\kappa}
\def\alam{A_\lam}

\newcommand{\ba}{\begin{array}}
\newcommand{\ea}{\end{array}}
\newcommand{\be}{\begin{equation}}
\newcommand{\ee}{\end{equation}}

\def\lsim{\mathrel{\raise.3ex\hbox{$<$\kern-.75em\lower1ex\hbox{$\sim$}}}} 
\def\gsim{\mathrel{\raise.3ex\hbox{$>$\kern-.75em\lower1ex\hbox{$\sim$}}}}

\def\beq{\begin{eqnarray}}
\def\eeq{\end{eqnarray}}
\def\bea{\begin{eqnarray}}
\def\eea{\end{eqnarray}}

\def\gev{\, {\rm GeV}}

\def\omhsq{\Omega_{\chi^0} h^2}
\def\tanb{\tan \beta}
\def\anti{\overline}
\def\hi{h_1}
\def\ai{a_1}
\def\mhi{m_{\hi}}
\def\mai{m_{\ai}}

\def\aii{a_2}

\def\hiii{h_3}

\def\mchi{m_{\chi^0}}

\begin{document}


\title{CoGeNT, DAMA, and Light Neutralino Dark Matter}
\author{Alexander V.~Belikov$^1$, John F.~Gunion$^{2}$, Dan Hooper$^{3,4}$, and Tim M.P. Tait$^5$}

\address{
${^1}$ Department of Physics, The University of Chicago, Chicago, IL 60637\\$^2$Department of Physics, University of California, Davis, CA 95616 \\ $^3$ Center For Particle Astrophysics, Fermi National Accelerator Laboratory, Batavia, IL 60510\\ $^4$ Department of Astronomy and Astrophysics, University of Chicago, Chicago, IL 60637\\ $^5$ Department of Physics and Astronomy, University of California, Irvine CA 92697}

\begin{abstract}

  Recent observations by the CoGeNT collaboration (as well as long
  standing observations by DAMA/LIBRA) suggest the presence of a $\sim
  5$-10 GeV dark matter particle with a somewhat large elastic
  scattering cross section with nucleons ($\sigma\sim 10^{-40}$
  cm$^2$).  Within the context of the minimal supersymmetric standard
  model (MSSM), neutralinos in this mass range are not able to possess
  such large cross sections, and would be overproduced in the early
  universe. Simple extensions of the MSSM, however, can easily
  accommodate these observations.  In particular, the extension of the
  MSSM by a chiral singlet superfield allows for the possibility that
  the dark matter is made up of a light singlino that interacts with
  nucleons largely through the exchange of a fairly light ($\sim$30-70
  GeV) singlet-like scalar higgs, $\hi$. Such a scenario is consistent
  with all current collider constraints and can generate the signals
  reported by CoGeNT and DAMA/LIBRA. Furthermore, there is a generic
  limit of the extended model in which there is a singlet-like
  pseudoscalar higgs, $\ai$, with $\mai\sim \mhi$ and in which the
  $\chi^0\chi^0$ and $b\anti b, s\anti s$ coupling magnitudes of the
  $\hi$ and $\ai$ are very similar.  In this case, the thermal relic
  abundance is automatically consistent with the measured density of
  dark matter if $\mchi$ is sufficiently small that $\chi^0\chi^0\to b\anti
  b$ is forbidden.

\end{abstract}

\pacs{95.35.+d, 12.60.Jv, 14.80.Da; FERMILAB-PUB-10-316-A}

\maketitle



Recently, the CoGeNT collaboration has reported the detection of very low energy events which cannot be accounted for with known backgrounds~\cite{cogentnew}. It has been shown than it is possible to interpret these events as the elastic scattering of a light dark matter particle ($m\sim 5-10$ GeV) with a cross section on the order of $\sim$$10^{-40}$ cm$^2$~\cite{cogentnew,liam,Kopp:2009qt,Chang:2010yk,collar}. Intriguingly, the range of masses and cross sections implied by CoGeNT is not very far from the region required to explain the annual modulation observed by the DAMA/LIBRA collaboration~\cite{DAMAnew,petriello,annualmodulation}. While null results from liquid XENON-based experiments~\cite{nullxenon} and CDMS~\cite{nullcdms} somewhat constrain dark matter interpretations of the CoGeNT and DAMA/LIBRA signals, uncertainties in the dark matter velocity distribution, as well as the various experiments' energy scale calibrations and quenching factors (and/or scintillation efficiencies) could potentially lead to a consistent interpretation~\cite{collar,uncertain}.

Since the announcement of the CoGeNT result, a number of groups have begun to explore 
the dark matter phenomenology of this 
signal~\cite{Andreas:2010dz,Essig:2010ye,Graham:2010ca,collar}.
Within the context of the Minimal Supersymmetric Standard Model (MSSM), dark matter 
explanations for the CoGeNT/DAMA signals face considerable challenges. The range 
of elastic scattering cross sections predicted for neutralinos falls more than an order of 
magnitude short, even in the 
most optimal regions of parameter space~\cite{Kuflik:2010ah,Feldman:2010ke}. While this could 
plausibly be reconciled by adopting a significantly higher local density of dark matter (the required cross section scales inversely with the local dark matter density), the relic 
abundance of very light (5-10 GeV) neutralinos in the MSSM is also predicted to be well above the 
measured cosmological dark matter density~\cite{Feldman:2010ke} (for earlier work on light neutralino dark matter in the MSSM, see Ref.~\cite{lightLSP}). Thus, even in
optimistic regions of the MSSM parameter space, it is very difficult to accommodate the 
observations of CoGeNT and DAMA/LIBRA. 

To increase the elastic scattering cross section and reduce the thermal relic abundance of 
neutralino dark matter, one could consider a combination of larger couplings or lower masses for 
the particles exchanged.  However, for very light neutralinos (whose scattering
is typically dominated by scalar higgs bosons) this prospect is constrained by
Tevatron and LEP II data which require the masses of MSSM
higgs bosons to lie above $\gsim 90$ GeV. However, the constraints
need not apply in supersymmetric models with extended higgs sectors \cite{mcelrath}. 
As we will show, in such scenarios it is possible for a 5-10 GeV neutralino to produce the 
observed signal through the exchange of a light ($\sim$30-70 GeV) scalar higgs, 
while also generating the correct thermal relic abundance. 

Generically, the spin-independent elastic scattering cross section of 
dark matter with a nucleus is written:
\begin{equation}
\label{sig}
\sigma \approx \frac{4 m^2_{\rm DM} m^2_{N}}{\pi (m_{\rm DM}+m_N)^2} [Z f_p + (A-Z) f_n]^2,
\end{equation}
where $m_N$ is the mass of the target nucleus (of atomic number $Z$ and mass $A$),
and $m_{\rm DM}$ is the dark matter mass.  
$f_p$ and $f_n$ are the dark matter's couplings to protons and neutrons:
\begin{equation}
f_{p,n}=\sum_{q=u,d,s} f^{(p,n)}_{T_q} a_q \frac{m_{p,n}}{m_q} + \frac{2}{27} f^{(p,n)}_{TG} \sum_{q=c,b,t} a_q  \frac{m_{p,n}}{m_q},
\label{feqn}
\end{equation}
where $a_q$ are the dark matter's couplings to quarks (in the
Lagrangian) and
$f^{(p,n)}_{T_q}$, $f^{(p,n)}_{TG}$ are hadronic matrix elements~\cite{nuc}.
An appropriate nuclear form factor accounts for the effects of finite momentum transfer.
We use the values $f^{(p)}_{T_u} = 0.020 \pm 0.004$, $f^{(p)}_{T_d} = 0.026 \pm 0.005$, $f^{(p)}_{T_s} = 0.118 \pm 0.062$, and $f^{(p)}_{TG} = 0.84$~\cite{scatteraq}.
The uncertainties in determination of hadronic matrix elements, which are due to the uncertainty in determination of $\sigma_{\pi N}$, might reach approximately a factor of two and therefore result in a factor of two uncertainty in the dark matter couplings. 



For light MSSM neutralinos, the neutralino-quark coupling 
is dominated by scalar higgs exchange
(contributions from squark exchange are typically negligible). 
For down-type quarks, this coupling is~\cite{scatteraq}:
\begin{eqnarray}
\label{ad}
\frac{a_d}{m_d}  &=&  \mbox{} \frac{g_2}{4 m_{W} \cos\beta} [-g_1 N_{11}+g_2 N_{12}]  \\
&\times& \bigg[\bigg(\frac{N_{13}c^2_{\alpha} - N_{14} c_{\alpha} s_{\alpha}}{m^2_H}\bigg)+\bigg(\frac{N_{13} s^2_{\alpha} + N_{14} c_{\alpha} s_{\alpha}}{m^2_h}\bigg)\bigg] \nonumber,
\end{eqnarray}
where the $N_{1i}$'s denote the composition of the lightest neutralino 
($\chi^0_1 = N_{11}\tilde{B}+N_{12} \tilde{W}^3+N_{13} \tilde{H}_d +N_{14} \tilde{H}_u$), 
and $s_{\alpha}$ and $c_{\alpha}$ denote the sine and cosine of $\alpha$, which relate
the scalar mass and gauge eigenstates.
The corresponding expression for up-type quarks is found by replacing 
$\cos \beta \leftrightarrow \sin \beta$ and $N_{14} \leftrightarrow N_{13}$.

The largest elastic scattering cross sections in the MSSM
arise in the case of large $\tan \beta$ and $\sin(\beta-\alpha)\sim 1$,
significant $N_{13}$, and relatively light $m_H$. In this limit, the lighter higgs, $h$, is approximately standard model-like and the heavier $H$ is approximately $H_d$, and one finds
\begin{equation}
\frac{a_d}{m_d}\approx \frac{-g_2 g_1 N_{13}N_{11} \tan \beta \, c^2_{\alpha}}{4 m_W m^2_H},
\end{equation}
which in turn yields $\sigma_{\chi^0 p,n} \approx$
%
\begin{equation}
1.8 \times 10^{-41} {\rm cm}^2 
\bigg(\frac{N^2_{13}}{0.103}\bigg) \bigg(\frac{\tan \beta}{50}\bigg)^2 
\bigg(\frac{90\, {\rm GeV}}{m_H}\bigg)^4  \bigg(\frac{c_{\alpha}}{1}\bigg)^4\,,
\label{mssmelastic}
\end{equation}
where the reference values for the Higgs mass, $\tanb$, $N_{13}^2$ and
$c_\alpha$ have been chosen in the most optimistic way as discussed
below.

The higgsino content of the lightest neutralino is constrained by the invisible width of the $Z$ as 
measured at LEP, $\Gamma^{\rm LEP}_{\rm inv} =499 \pm 1.5$ MeV. In contrast, the standard 
model prediction for this quantity is slightly (1.4$\sigma$) higher, 
$\Gamma^{\rm SM}_{\rm inv}=501.3 \pm 0.6$ MeV~\cite{zwidth}. Combining the measured and 
predicted values, we find a $2\sigma$ upper limit of $\Gamma_{Z\rightarrow \chi^0 \chi^0} < 1.9$ 
MeV. As $\Gamma_{Z\rightarrow \chi^0 \chi^0}$ scales with $[N^2_{13}-N^2_{14}]^2$, we can 
translate this result to a limit of $|N^2_{13}-N^2_{14}| < 0.103$. For moderately large values of 
$\tan \beta$, the two higgsino terms do not efficiently cancel, requiring $|N_{13}|^2 < 0.103$.  

$m_H$ and $\tan \beta$ are constrained by a number of measurements, including 
those of the rare decays
$t\rightarrow b H^+$, $B_s \rightarrow \mu^+ \mu^-$, $B^{\pm}
\rightarrow \tau \nu$, and direct limits on higgs production followed
by $A/H\rightarrow \tau^+ \tau^-$. While these limits vary somewhat
depending on the precise values of the MSSM parameters adopted, in
general they imply $\tan \beta \lsim 20-30$  for 
$m_H, m_A \sim 90-150$ GeV (the strongest limits
coming from the latest LHC results). Constraints from LEP II 
further require $m_{H,A} \gsim 90$ GeV. When these limits are taken
into account, we find that $\sigma_{\chi^0 p,n} \lsim 10^{-41}$
cm$^2$~\cite{Kuflik:2010ah,Feldman:2010ke}, which falls short of that
implied by the CoGeNT and DAMA/LIBRA signal by about an order of
magnitude. 

Furthermore, light neutralinos in the MSSM are inevitably predicted to
freeze out with a thermal relic abundance in excess of the measured
dark matter density.  Indeed, there is a kind of inverse relation
whereby for $m_H\sim m_A$ near 90 GeV, appropriate $\omhsq$ from
annihilation via the $A$ to $b\anti b$ and $\tau^+\tau^-$ is roughly
proportional to the inverse of $\sigma_{\chi^0 p,n} $ and only falls
to a value of order $\omhsq\sim 0.1$ when $\sigma_{\chi^0 p,n} $ is of
order the maximal value indicated above.  Given that current
experimental constraints in the MSSM context do not allow one to
achieve this maximal value, the MSSM inevitably leads to too large a
value for $\omhsq$. Increasing $m_A$ or $m_H$ worsens the situation.


To increase the cross section beyond the range allowed in the MSSM, an
obvious direction is to consider models with lighter higgs bosons. As
the cross section scales with the inverse of the fourth power of the
exchanged higgs mass, even modest reductions in $m_H$ could increase
the cross section to the levels required.  As an example of a
framework in which light higgs bosons are possible, we extend the MSSM
by a chiral singlet superfield $\hat{S}$, containing two neutral
scalars $H_S$ and $A_S$ and a Majorana fermion $\tilde{S}$.  The
theory is described by superpotential \cite{Hooper:2009gm}
\begin{equation}
\frac{1}{2} \mu_S \hat{S}^2 + \mu \hat{H}_u \hat{H}_d + \lambda \hat{S} \hat{H}_u \hat{H}_d
+ \frac{1}{3} \kappa \hat{S}^3 ~,
\label{eq:W}
\end{equation}
(along with the MSSM Yukawa interactions)
and soft Lagrangian
\begin{eqnarray}
{\cal L}_{soft} & = & v_S^3 S + 
B_{\mu} H_u H_d+\frac{1}{2} m_S^2 |S|^2 + 
\frac{1}{2} B_S S^2 \nonumber \\ & &
+ \lambda A_\lambda S H_u H_d +  \frac{1}{3}\kappa A_\kappa S^3 + H.c.
\label{eq:Lsoft}
\end{eqnarray}
(along with the MSSM $A$-terms).
Specific implementations of such a singlet typically involve a subset of these terms. For example, in the Next-to-Minimal Supersymmetric Standard Model (NMSSM)~\cite{nmssm}, a $Z_3$ symmetry is imposed which only allows the terms involving 
$\lambda$, $\kappa$, $A_{\lambda}$ and $A_{\kappa}$. Here, we do not tie ourselves to this 
particular model, but instead consider the full range of terms as described in 
Eqns.~(\ref{eq:W}) and~(\ref{eq:Lsoft}), which we refer to as the
Extended Next-to-Minimal supersymmetric Standard Model (ENMSSM).
	
The tree-level neutralino mass matrix 
in the $\tilde{B}, \tilde{W}^3, \tilde{H}_u, \tilde{H}_d, \tilde{S}$ basis is
\beq
{\cal M}_{\wtil\chi^0} =
\left( \ba{ccccc}
M_1 & 0 & \frac{g_1 v_u}{\sqrt{2}} & -\frac{g_1 v_d}{\sqrt{2}} & 0 \\
0 & M_2 & -\frac{g_2 v_u}{\sqrt{2}} & \frac{g_2 v_d}{\sqrt{2}} & 0 \\
\frac{g_1 v_u}{\sqrt{2}} & -\frac{g_2 v_u}{\sqrt{2}} & 0 & -\mu-\lambda s & -\lam v_d \\
 -\frac{g_1 v_d}{\sqrt{2}} &  \frac{g_2 v_d}{\sqrt{2}} & -\mu-\lambda s & 0 & -\lam v_u \\
0 & 0 & -\lam v_d & -\lam v_u & 2 \kap s+\mu_S
\ea \right) , \eeq
where $v_u$ and $v_d$ are the up- and down-type higgs vevs and $s$ is
the vev of the singlet higgs. A light ($\lsim 10$~GeV) neutralino
consistent with LEP II chargino searches must be either mostly
$\tilde{B}$ or $\tilde{S}$.  The $\tilde{B}$ does not couple to a
mostly singlet higgs, and that case is thus similar to the MSSM.  From
here on, we focus on the case where the lightest neutralino is mostly
$\tilde{S}$.  One might imagine that the strict NMSSM would allow
sufficient flexibility.  It has been shown~\cite{us} that for the bulk
of the NMSSM, after imposing LEP and $B$-physics constraints the
lightest neutralino is always bino-like and elastic cross sections as
large as required by CoGeNT and DAMA/LIBRA are not possible.
Nonetheless they are ``only'' a factor of 10 too small (whereas in
Ref.~\cite{Das:2010ww} the largest cross section found was a factor of
100 too small). An exception to this conclusion was identified in
Ref.~\cite{wagner}, in which a very light ($\sim$GeV) singlet scalar
higgs is able to generate the very large elastic scattering cross
section required by CoGeNT and DAMA/LIBRA within the context of the
NMSSM. These scenarios require a considerable degree of fine-tuning of
the parameters. As we show, in the ENMSSM, it is possible to find less
fine-tuned scenarios with large elastic cross section and correct
relic density when the light higgs has mass $\gsim 30$ GeV even when
$\tanb$ is small enough to evade the most recent LHC limits on Higgs
bosons with enhanced $b\anti b$ couplings proportional to $\tanb$.



We proceed by engineering the lightest neutralino to be mostly
$\tilde{S}$, together with a light higgs that is predominantly
singlet.  The lightest neutralino will naturally be predominantly
singlino provided the quantity $|2 \kappa s + \mu_S|$ is much smaller
than $|\mu+\lambda s|$, $M_1$ and $M_2$. For example, for $\kappa \sim
0.45$, $s\sim 2$ GeV, $\mu_S\approx 0$, $\lambda \sim 0.01$, $\tan
\beta\sim 15$, large $M_1$, large $M_2$, and $\mu\sim$180 GeV, we find
that the lightest neutralino is singlino-like ($N^2_{15}=0.99$) with a
mass of approximately 5 GeV.

The conditions under which the lightest higgs, $h_1$, is mostly
singlet are somewhat more complicated.  A simple limit which leads to
desired phenomena can be obtained for small $\lambda$.  In the limit
$\lambda \rightarrow 0$, but keeping $\lam \alam$ of moderate size,
the singlet decouples from the MSSM (which has standard higgses), and
has a mass determined primarily by the $\lam\alam v^2 \sin 2\beta
/(2s)$ singlet-singlet entry in the mass-squared matrix, with weak
dependence on $B_S$, $m^2_S$, $\mu_S$, $\kappa$, and $A_\kappa$.
Then, provided the soft terms for the singlet are sufficiently small,
and $\lambda v$ is much smaller than $m_h$, the singlet represents a
perturbation on MSSM higgs phenomenology, with a light singlet state
mixed with the MSSM to a degree controlled by $\lambda$.  In this
limit, the light CP odd higgs will also be predominantly singlet, with
mass and mixings to the MSSM $A$ that are proportional to $\lam\alam
v^2 \sin 2\beta/(2s)$ {\it with the same coefficients as for the
  $\hi$}. Effectively, the $\hi$ and $\ai$ combine to form a single
complex nearly singlet scalar state.  Through mixing with the MSSM
higgses, both the $\hi$ and $\ai$ have couplings proportional to the
usual MSSM interactions, but reduced by the small amount of mixing.
For the parameters listed earlier, along with $A_\kappa\sim 33$ GeV,
$A_\lambda\sim 3400$ GeV, $B_S\sim 0$, $m_S^2\sim 0$ one finds
$\mhi\sim 40\gev$, $\mai\sim 35\gev$ with $|F_s(\hi)|^2\sim
|F_s(\ai)|^2\sim 0.65$ and $|F_d(\hi)|^2\sim |F_d(\ai)|^2\sim 0.35$,
where $F_s(\hi)$ and $F_d(\hi)$ are the singlet and $H_d^0$ components
(at the amplitude level) of the $\hi$ and $F_s(\ai)$ and $F_d(\ai)$
are the singlet and $A_d^0$ components of the $\ai$. To reemphasize,
in the scenarios we consider the $\ai$ and $\hi$ are close in mass and
have $F_s$ and $F_d$ components that are very similar.

The singlino coupling to down-type quarks via $\hi$ exchange at Lagrangian
level is given by
\begin{equation}
\frac{a_d}{m_d} =  \frac{g_2 \kappa N^2_{15}  F_s(\hi) F_d(\hi)}{2\sqrt
  2 m_W m^2_{h_1}\cos\beta}
\end{equation}
 For $\hi$ exchange only, the resulting spin-independent cross section is
\begin{eqnarray}
\sigma_{\chi^0 p,n} &\approx& 3.3 \times 10^{-40} \, {\rm cm}^2 \label{eq:sigma} 
\bigg(\frac{\kappa}{0.4}\bigg)^2 \, \bigg(\frac{\tan
  \beta}{15}\bigg)^2\cr &\times& \bigg(\frac{40\, {\rm GeV}}{m_{h_1}}\bigg)^4
\bigg(\frac{|F_s(\hi)|^2}{0.65}\bigg)
\bigg(\frac{|F_d(\hi)|^2}{0.35}\bigg), 
\end{eqnarray}
which is of order the value required by CoGeNT and
DAMA/LIBRA. Furthermore, the mostly singlet nature ($|F_s(\hi)|^2 \sim
0.65$) of the $h_1$ easily allows it to evade the constraints from LEP
II and the Tevatron, as we discuss below.  The fact that moderately
large $|F_d(\hi)|^2$ is required argues that this scenario will be
difficult to arrange in the (MSSM) decoupling regime of $m_{a_2} \gg
m_{h_2}$, implying that all of the mostly MSSM higgses are likely to
have masses only slightly above the LEP II limit. In fact, for the
parameters listed above, $m_{h_2}\sim 109$ GeV (but escapes LEP limits
since it has substantial singlet component, $|F_s(h_2)|^2\sim 0.28$),
$m_{a_2}\sim 111$ GeV and $m_{h_3}\sim m_{h^+}\sim 125$ GeV.  After
including $h_2$ and $h_3$ in the computation of $\sigma_{\chi^0 p,n}$
the coefficient of 3.3 is reduced to about 2.2 in
Eq.~(\ref{eq:sigma}) due to some partial cancellation at the amplitude level.

Collider constraints on the $h_1$ and $h_2$ are largely evaded for
sufficiently small $|F_s(\hi)|$ and $|F_s(h_2)|$.  LEP II places
constraints through production of $Z h_i$ (for the scalars) and pair
production of $a_i h_j$ for the pseudoscalars.  The heavier mass
eigenstates ($h_3$, and $a_2$) look like their MSSM counterparts, with
their couplings slightly reduced by a small singlet component.  Thus,
provided they represent a viable point of MSSM parameter space, they
will be allowed here as well.  The light (mostly singlet) $h_1$ and
$a_1$ must have small enough $Z$-$Z$-$h_1$ and $Z$-$h_i$-$a_1$
interactions to be consistent with existing searches.  In the
scenarios considered here, the coupling of the $\hi$ to $WW,ZZ$
(relative to the SM coupling), denoted $C_V(\hi)$, is very small,
$|C_V(\hi)|<0.1$, which easily allows $m_{h_1} \sim 40-50$ GeV to be
consistent with LEP limits on $Z\hi$~\cite{Schael:2006cr}.  Small
$|C_V(\hi)|$ arises in the limit $A_\lambda \gsim \mu$.  The
$Z$-$h_1$-$a_1$ coupling is similarly suppressed implying that LEP
limits on $h_1 a_1$ pair production are easily evaded.  Pair
production of $\ai$ together with the mostly SM-like light higgs $h_2$
is sufficiently suppressed by the largely singlet nature of the $\ai$.
The Tevatron can produce scalars and pseudoscalars through the
reaction $b g \rightarrow b +h,a$, where the $h$ or $a$ can decay into
either $b \bar{b}$ or $\tau^+ \tau^-$ pairs. In our scenarios, the
strongest constraints arise for $h=\hiii$ and $a=\aii$ since they are
fairly light and mainly non-singlet and have $b\anti b$ couplings that
are enhanced at large $\tanb$.  Null LHC searches with $L=1$ fb$^{-1}$
of accumulated luminosity at CMS require $\tan \beta \lsim 25$ for a
doublet-like $\hiii$ or $\aii$ with (non-degenerate) masses of order 100
GeV~\cite{hig11009}.
A similar limit on $\tan \beta$ can be obtained from the null search
for $t \rightarrow H^+ b$~\cite{hig11008}.
Taken all together, the central parameters of Eqn.~(\ref{eq:sigma})
are on the border of a number of higgs searches, and are thus being
tested by end-phase Tevatron and early LHC running.

The thermal relic density of neutralinos is determined by the
annihilation cross section and neutralino mass. In the $m_{\chi^0}$
range we are considering here, the potentially important annihilation
channels are to $b\bar{b}$ or $\tau^+ \tau^-$ through the $s$-channel
exchange of Higgs bosons. Given that $\mhi\sim \mai$ and that the $F_s$
and $F_d$ components of the $\hi$ and $\ai$ are similar, the CP-odd
$\ai$ is dominant, annihilation via the CP-even $\hi$ being p-wave
($v^2$) suppressed.\footnote{This differs from the scenario of
  \cite{wagner} in which the $\ai$ is highly singlet and annihilation
  is dominated by the $\hi$.} The annihilation cross section for the
$b\anti b$ final state that results from $s$-channel exchange of the
$\ai$ is given by
\begin{eqnarray}
\sigma v &=& 
\frac{N_c g^2_2 \kappa^2 m^2_b |F_s(\ai)|^2 |F_d(\ai)|^2}{4 \pi m^2_W
  \cos^2\beta}\cr
&&\quad \qquad\times
\frac{m^2_{\chi^0} (1-m^2_b/m^2_{\chi^0})^{1/2} }
{(4 m^2_{\chi^0}-m^2_{a_1})^2 + m^2_{a_1} \Gamma^2_{a_1}},~~~
\end{eqnarray}
%
where $v$ is relative velocity between the annihilating neutralinos,
$N_c = 3$ is a color factor and $\Gamma_{a_1}$ is the width of the
exchanged higgs. The annihilation cross section into $\tau^+ \tau^-$
is obtained by replacing $m_b \rightarrow m_{\tau}$ and $N_c
\rightarrow 1$. The thermal relic abundance of
neutralinos is obtained as
\begin{equation}
\Omega_{\chi^0} h^2 \approx \frac{10^9}{M_{\rm Pl}}\frac{m_{\chi^0}}{T_{\rm FO} \sqrt{g_{\star}}}
\frac{1}{\langle \sigma_{\chi^0 \chi^0} v \rangle},
\end{equation}
where $g_{\star}$ is the number of relativistic degrees of freedom
available at freeze-out, $\langle \sigma_{\chi^0 \chi^0} v \rangle$ is
the thermally averaged annihilation cross section at freeze-out, and
$T_{\rm FO}$ is the temperature at which freeze-out occurs.
%
%
In the scenarios considered here, one finds that the annihilation rate
is too large if the $\ai\to b\anti b$ channel is open. As a result,
consistency with the observed relic density is only ``automatically''
obtained if $\mchi<m_b(pole)\simeq 5.28$ GeV. 

For the range of masses and cross sections considered here, we find 
$m_{\chi^0}/T_{\rm FO}\approx 20$, leading to a thermal relic
abundance from $\ai\to \tau^+\tau^-$ of
\begin{eqnarray}
\Omega_{\chi^0} h^2 
\approx 0.15 \,  \bigg(\frac{0.4}{\kappa}\bigg)^2 \bigg(\frac{15}{\tan \beta}\bigg)^2 \bigg(\frac{m_{a_1}}{35 \, {\rm GeV}}\bigg)^4 \nonumber \\
\times \bigg(\frac{5 \, {\rm GeV}}{m_{\chi^0}}\bigg)^2  \bigg(\frac{0.65}{|F_s(\ai)|^2}\bigg) \bigg(\frac{0.35}{|F_d(\ai)|^2}\bigg),
\end{eqnarray}
applicable so long as $\mchi$ is small enough that the $b\anti b$
channel is not open.  After including other exchanges, one obtains a
value of $\Omega_{\chi^0} h^2$ that is consistent with the measured
dark matter density, $\Omega_{\rm CDM} h^2 = 0.1131 \pm
0.0042$~\cite{wmap}.

Thus, the desired relic density automatically results at low $\mchi$
once the relevant combination of couplings and higgs masses are set to
accommodate CoGeNT and DAMA/LIBRA. 
This confluence of parameter space is peculiar to models with
scalar/pseudoscalar exchange~\cite{modelindependent,liam,portal}.  For
example, a Dirac fermion or a scalar with vector interactions will
either overproduce the CoGeNT and DAMA/LIBRA rates or will predict a
thermal relic density in excess of the measured dark matter abundance.

This scenario also has interesting implications for the indirect
detection of dark matter.  In particular, as the dark matter
annihilation rate in any given region scales with the inverse of the
square of the dark matter mass, the light neutralino we are
considering could, in principle, lead to enhanced fluxes of various
annihilation products~\cite{Bottino:2004qi}. Quantitatively, the spectrum of gamma-rays from dark matter annihilations can be written as
\begin{equation}
\Phi_{\gamma}(E_{\gamma},\psi) =  \frac{d N_{\gamma}}{d E_{\gamma}} \frac{\sigma v}{8\pi m^2_{\chi^0}} \int_{\rm{los}} \rho^2(r) dl,
\label{flux1}
\end{equation}
where $\sigma v$ is the dark matter annihilation cross section multiplied by the relative velocity of the two neutralinos, $\psi$ is the angle observed relative to the direction of the Galactic Center, $\rho(r)$ is the dark matter density as a function of distance to the Galactic Center, $d N_{\gamma}/d E_{\gamma}$ is the gamma ray spectrum generated per annihilation, and the integral is performed over the line-of-sight. For a neutralino with $\mchi\sim 5$ GeV and with an annihilation cross section of $\sigma v \approx 3 \times 10^{-26}$ cm$^3$/s to $\tau^+ \tau^-$, the annihilation rate in the Galactic Center (assuming a NFW halo distribution) is predicted to lead to a flux of gamma-rays above 1 GeV of $\approx$ 2.9 cm$^{-2}$~yr$^{-1}$ from the inner degree of our galaxy, corresponding to thousands of events per year observed by the Fermi Gamma Ray Space Telescope (FGST). Furthermore, the flux and spectral shape of the gamma-ray emission observed by the FGST from this region of the sky is quite similar to that predicted from dark matter annihilations~\cite{Hooper:2010mq}. FGST's observations of dwarf spheroidal galaxies~\cite{Abdo:2010ex} are also potentially sensitive to a dark matter particle with these characteristics.

The prediction that the dark matter annihilates primarily to $\tau^+\tau^-$ also insures that our dark matter candidate will not violate constraints from cosmic ray antiproton measurements, as set by the PAMELA experiment~\cite{Bottino:2004qi,Bottino:2005xy,Adriani:2010rc}. If gamma ray searches for the products of dark matter annihilations were in the future to constrain the low-velocity annihilation cross section to be well below the value of $\sigma v \sim 3 \times 10^{-26}$ cm$^3$/s, this would rule out the present model as well as the possibility that the dark matter is a scalar with scalar interactions~\cite{modelindependent,liam,Arina:2010rb}.

Another indirect detection mode with decent prospects 
is to search for energetic neutrinos produced through the
capture and annihilation of dark matter in the core of the Sun. The Sun is predicted to capture dark matter particles at a rate given by

\begin{eqnarray}
C^{\odot} &\simeq& 3.5 \times 10^{24} \, \mathrm{s}^{-1} 
\left( \frac{\rho_{\chi^0}}{0.4\, \mathrm{GeV}/\mathrm{cm}^3} \right) 
\left( \frac{270\, \mathrm{km/s}}{\bar{v}} \right)  
\left( \frac{5 \, \mathrm{GeV}}{m_{\chi^0}} \right)  \nonumber \\
&\times& \bigg[ \bigg(\frac{\sigma_{\mathrm{H}}}{10^{-40}\, {\rm cm}^2}\bigg) +  1.1 \bigg(\frac{\sigma_{\mathrm{He}}}{16 \times 10^{-40}\, {\rm cm}^2}\bigg)\bigg],
\label{capture}
\end{eqnarray}
where $\rho_{\chi^0}$ is the local dark-matter density, $\bar{v}$ is the local root-mean-square velocity of halo dark-matter particles, and $\sigma_{\mathrm{H}}$ and $\sigma_{\mathrm{He}}$ are the elastic scattering cross sections of the WIMP with hydrogen and helium nuclei, respectively. In the model under consideration, the elastic scattering cross section of the neutralino is sufficiently large that the processes of capture and annihilation quickly reach equilibrium in the Sun, removing any dependence on the neutralino's annihilation cross section. The high capture rate in this model is predicted to produce a sizable flux of GeV-scale neutrinos, comparable to the constraints currently placed by
Super-Kamiokande~\cite{Kappl:2011kz,liam,neutrino}. In particular, for $\mchi\sim 5~\mathrm{GeV}$ and annihilation entirely to $\tau^+ \tau^-$, Super-Kamiokande data can be used to constrain $\sigma_{\chi^0 p} \lsim 6\times10^{-41} \mathrm{cm}^2$ \cite{liam}, for reasonable astrophysical assumptions. Larger volume neutrino experiments such as IceCube have energy thresholds which are too high to observe the annihilation products of such light dark matter particles.

In summary, we have considered the possibility that neutralino dark
matter is responsible for the signals reported by the CoGeNT and
DAMA/LIBRA collaborations. Although, the elastic scattering cross
section of neutralinos with nuclei in the MSSM is too small to account
for these observations, the same conclusion is not necessarily reached
in extended supersymmetric models. In particular, we have discussed
models in which the MSSM is extended by a chiral singlet
superfield. In such a model, a light singlino-like neutralino, which
interacts with nuclei through the exchange of a largely singlet-like,
scalar higgs, can possess an elastic scattering cross section capable
of generating the observations reported by CoGeNT and
DAMA/LIBRA. Furthermore, the scenarios considered automatically lead
to a thermal relic abundance of neutralinos consistent with the
observed density of dark matter for $\mchi\lsim 5$ GeV.

After the completion of this project, the CoGeNT collaboration reported the detection of an annual modulation of their rate at the level of $2.8\sigma$~\cite{CoGeNTannual}. This result provides further motivation for the type of model considered in this paper.

\section*{Acknowledgements} AB and DH are supported by the US
Department of Energy, including grant DE-FG02-95ER40896, and by NASA
grant NAG5-10842. JFG is supported by US DOE grant
DE-FG03-91ER40674.  TT is supported by NSF grant PHY-0970171
and acknowledges the hospitality of the SLAC theory group.


\begin{thebibliography}{9}


\bibitem{cogentnew}
C.~E.~Aalseth {\it et al.} [The CoGeNT Collaboration],
arXiv:1002.4703 [astro-ph.CO].

\bibitem{liam}
  A.~L.~Fitzpatrick, D.~Hooper and K.~M.~Zurek,
  arXiv:1003.0014 [hep-ph].

\bibitem{Kopp:2009qt}
  J.~Kopp, T.~Schwetz and J.~Zupan,
  JCAP {\bf 1002}, 014 (2010)
  [arXiv:0912.4264 [hep-ph]].

\bibitem{Chang:2010yk}
  S.~Chang, J.~Liu, A.~Pierce, N.~Weiner and I.~Yavin,
  arXiv:1004.0697 [hep-ph].

\bibitem{collar}
  D.~Hooper, J.~I.~Collar, J.~Hall and D.~McKinsey,
  [arXiv:1007.1005 [hep-ph]].

\bibitem{DAMAnew}
  R.~Bernabei {\it et al.},
  arXiv:1002.1028 [astro-ph.GA].

\bibitem{petriello}
  F.~Petriello and K.~M.~Zurek,
  JHEP {\bf 0809}, 047 (2008)
  [arXiv:0806.3989 [hep-ph]];
A.~Bottino, F.~Donato, N.~Fornengo and S.~Scopel, 
Phys.\ Rev.\ {\bf D69}, 037302 (2004).



\bibitem{annualmodulation}
  A.~K.~Drukier, K.~Freese, D.~N.~Spergel,
  Phys.\ Rev.\  {\bf D33}, 3495-3508 (1986).


\bibitem{nullxenon}
 J.~Angle {\it et al.}  [XENON Collaboration],
  Phys.\ Rev.\ Lett.\  {\bf 100}, 021303 (2008)
  [arXiv:0706.0039 [astro-ph]];
  C.~Savage, G.~Gelmini, P.~Gondolo {\it et al.},
  [arXiv:1006.0972 [astro-ph.CO]].

\bibitem{nullcdms}
  Z.~Ahmed {\it et al.} [ CDMS-II Collaboration ],
  [arXiv:1011.2482 [astro-ph.CO]];
  D.~S.~Akerib {\it et al.} [ CDMS Collaboration ],
  Phys.\ Rev.\  {\bf D82}, 122004 (2010).
  [arXiv:1010.4290 [astro-ph.CO]].


\bibitem{uncertain}
  J.~I.~Collar,
  [arXiv:1010.5187 [astro-ph.IM]];
  J.~I.~Collar, D.~N.~McKinsey,
  [arXiv:1005.0838 [astro-ph.CO]]; [arXiv:1005.3723 [astro-ph.CO]];
  P.~Sorensen, J.~Angle, E.~Aprile {\it et al.},
  [arXiv:1011.6439 [astro-ph.IM]];
J.~Collar, in preparation.


\bibitem{Andreas:2010dz}
  S.~Andreas, C.~Arina, T.~Hambye, F.~S.~Ling and M.~H.~G.~Tytgat,
  arXiv:1003.2595 [hep-ph].

\bibitem{Essig:2010ye}
  R.~Essig, J.~Kaplan, P.~Schuster and N.~Toro,
  arXiv:1004.0691 [hep-ph].

\bibitem{Graham:2010ca}
  P.~W.~Graham, R.~Harnik, S.~Rajendran and P.~Saraswat,
  arXiv:1004.0937 [hep-ph].






\bibitem{Kuflik:2010ah}
  E.~Kuflik, A.~Pierce and K.~M.~Zurek,
  arXiv:1003.0682 [hep-ph].

\bibitem{Feldman:2010ke}
  D.~Feldman, Z.~Liu and P.~Nath,
  arXiv:1003.0437 [hep-ph].
  

\bibitem{lightLSP}
  D.~Hooper and T.~Plehn,
  Phys.\ Lett.\  B {\bf 562}, 18 (2003)
  [arXiv:hep-ph/0212226];
  A.~Bottino, N.~Fornengo and S.~Scopel,
	Phys.\ Rev.\  D {\bf 67}, 063519 (2003)
	[arXiv:hep-ph/0212379],
	A.~Bottino, F.~Donato, N.~Fornengo and A.~Scopel,
	Phys.\ Rev.\  D {\bf 68}, 043506 (2003)
	[arXiv:hep-ph/0304080],
	A.~Bottino, F.~Donato, N.~Fornengo and A.~Scopel,
	Phys.\ Rev.\  D {\bf 69}, 037302 (2004)
	[arXiv:hep-ph/0307303],
	D.G.~Cerdeno and C.~Munoz,
	JHEP {\bf 10}, 15 (2004)
	[arXiv:hep-ph/0405057].

\bibitem{mcelrath}
  J.~F.~Gunion, D.~Hooper and B.~McElrath,
  Phys.\ Rev.\  D {\bf 73}, 015011 (2006)
  [arXiv:hep-ph/0509024];
see also K.~J.~Bae, H.~D.~Kim, S.~Shin,
  [arXiv:1005.5131 [hep-ph]].

\bibitem{nuc}
  A.~Bottino, F.~Donato, N.~Fornengo and S.~Scopel,
  Astropart.\ Phys.\  {\bf 18}, 205 (2002)
  [arXiv:hep-ph/0111229];
 Astropart.\ Phys.\  {\bf 13}, 215 (2000)
  [arXiv:hep-ph/9909228];
  J.~R.~Ellis, K.~A.~Olive, Y.~Santoso and V.~C.~Spanos,
  Phys.\ Rev.\ D {\bf 71}, 095007 (2005)
  [arXiv:hep-ph/0502001];
  J.~Giedt, A.~W.~Thomas and R.~D.~Young,
  arXiv:0907.4177 [hep-ph].
	J.~Ellis, K.~A.~Olive and C.~Savage,
  Phys.\ Rev.\ D {\bf 77}, 065026 (2008)
  [arXiv:hep-ph/0801.3656];
	
\bibitem{scatteraq}
G.~B.~Gelmini, P.~Gondolo and E.~Roulet,
Nucl.\ Phys.\ B {\bf 351}, 623 (1991);
M.~Srednicki and R.~Watkins,
Phys.\ Lett.\ B {\bf 225}, 140 (1989);
M.~Drees and M.~Nojiri,
Phys.\ Rev.\ D {\bf 48}, 3483 (1993)
[arXiv:hep-ph/9307208];-
M.~Drees and M.~M.~Nojiri,
Phys.\ Rev.\ D {\bf 47}, 4226 (1993)
[arXiv:hep-ph/9210272];
J.~R.~Ellis, A.~Ferstl and K.~A.~Olive, 
Phys.~Lett.~B  481, (2000) 304,
[arXiv:hep-ph/0001005].



\bibitem{zwidth}
C.~Amsler {\it et al}. (Particle Data Group), 
Phys.\ Lett.\  B {\bf 667}, 1 (2008).




\bibitem{Hooper:2009gm}
  D.~Hooper and T.~M.~P.~Tait,
  Phys.\ Rev.\  D {\bf 80}, 055028 (2009)
  [arXiv:0906.0362 [hep-ph]].


\bibitem{nmssm}
  J.~R.~Ellis, J.~F.~Gunion, H.~E.~Haber, L.~Roszkowski and F.~Zwirner,
  Phys.\ Rev.\  D {\bf 39}, 844 (1989);
  H.~P.~Nilles, M.~Srednicki and D.~Wyler,
  Phys.\ Lett.\  B {\bf 120}, 346 (1983);
  J.~E.~Kim and H.~P.~Nilles,
  Phys.\ Lett.\  B {\bf 138}, 150 (1984);
  M.~Drees,
  Int.\ J.\ Mod.\ Phys.\  A {\bf 4}, 3635 (1989).
  



\bibitem{us} 
  J.~F.~Gunion, A.~V.~Belikov, D.~Hooper,
  [arXiv:1009.2555 [hep-ph]].A.~Belikov, J.~Gunon and D.~Hooper, in preparation.

\bibitem{Das:2010ww}
  D.~Das, U.~Ellwanger,
[arXiv:1007.1151 [hep-ph]].




\bibitem{wagner}
  P.~Draper, T.~Liu, C.~E.~M.~Wagner {\it et al.},
  Phys.\ Rev.\  Lett. {\bf 106}, 121805 (2011)
  [arXiv:1009.3963 [hep-ph]].

\bibitem{Schael:2006cr}
  S.~Schael {\it et al.} [LEP Higgs Working Group],
  Eur.\ Phys.\ J.\  {\bf C47}, 547-587 (2006).
  [hep-ex/0602042].


\bibitem{hig11009}
[CMS Collaboration], CMS PAS HIG-11-009.



\bibitem{hig11008}
[CMS Collaboration], CMS PAS HIG-11-008.











\bibitem{wmap}
  E.~Komatsu {\it et al.}  [WMAP Collaboration],
  Astrophys.\ J.\ Suppl.\  {\bf 180}, 330 (2009)
  [arXiv:0803.0547 [astro-ph]].








  

  
  











 


  
  
  


  
 





  
  
  
  



\bibitem{modelindependent}
  M.~Beltran, D.~Hooper, E.~W.~Kolb and Z.~C.~Krusberg,
  Phys.\ Rev.\  D {\bf 80}, 043509 (2009)
  [arXiv:0808.3384 [hep-ph]].


\bibitem{portal}
  S.~Andreas, C.~Arina, T.~Hambye {\it et al.},
  [arXiv:1003.2595 [hep-ph]].









\bibitem{Bottino:2004qi}
  A.~Bottino, F.~Donato, N.~Fornengo, S.~Scopel,
  Phys.\ Rev.\  {\bf D70}, 015005 (2004).
  [hep-ph/0401186].

\bibitem{Abdo:2010ex}
  A.~A.~Abdo, M.~Ackermann, M.~Ajello, W.~B.~Atwood, L.~Baldini, J.~Ballet, G.~Barbiellini, D.~Bastieri {\it et al.},
  Astrophys.\ J.\  {\bf 712}, 147-158 (2010).
  [arXiv:1001.4531 [astro-ph.CO]]; see also results presented at the 2011 Fermi Symposium, \url{http://fermi.gsfc.nasa.gov/science/symposium/2011/}




\bibitem{Hooper:2010mq}
  D.~Hooper, L.~Goodenough,
  Phys.\ Lett.\  {\bf B697}, 412-428 (2011).
  [arXiv:1010.2752 [hep-ph]].




\bibitem{Bottino:2005xy}
  A.~Bottino, F.~Donato, N.~Fornengo, P.~Salati,
  Phys.\ Rev.\  {\bf D72}, 083518 (2005).
  [hep-ph/0507086].

\bibitem{Adriani:2010rc}
  O.~Adriani {\it et al.} [ PAMELA Collaboration ],
  Phys.\ Rev.\ Lett.\  {\bf 105}, 121101 (2010).
  [arXiv:1007.0821 [astro-ph.HE]].

\bibitem{Arina:2010rb}
  C.~Arina, M.~H.~G.~Tytgat,
	JCAP, {\bf 1101}, 011 (2011),
	[arXiv:1007.2765 [astro-ph.CO]].

\bibitem{neutrino}
  F.~Ferrer, L.~M.~Krauss and S.~Profumo,
  Phys.\ Rev.\  D {\bf 74}, 115007 (2006)
  [arXiv:hep-ph/0609257];
    V.~Niro, A.~Bottino, N.~Fornengo and S.~Scopel,
  Phys.\ Rev.\  D {\bf 80}, 095019 (2009)
  [arXiv:0909.2348 [hep-ph]].

\bibitem{Kappl:2011kz}
  R.~Kappl, M.~W.~Winkler,
  Nucl.\ Phys.\  {\bf B850}, 505-521 (2011).
  [arXiv:1104.0679 [hep-ph]].

\bibitem{CoGeNTannual}
		C.E.~Aalseth and others,
  	[arXiv:1106.0650 [astro-ph.CO]].






\end{thebibliography}
\end{document}